\documentclass[preprint]{elsarticle}

\usepackage{lineno,hyperref}
\usepackage{multirow}
\usepackage{siunitx}
\usepackage{graphicx}
\usepackage{subfig}
\usepackage{makecell}
\usepackage{array}
\usepackage{caption}
\usepackage{longtable}
\usepackage{float} 
\usepackage{amssymb}
\usepackage{amsmath}
\usepackage{subfig}
\usepackage{todonotes}
\usepackage[margin=1in]{geometry}
\newcolumntype{P}[1]{>{\centering\arraybackslash}p{#1}}

\usepackage{textcomp} 

\journal{International Journal of Heat and Mass Transfer}

\PassOptionsToPackage{compress}{natbib}
\begin{document}

\begin{frontmatter}

\title{Determination of Thermal Accommodation Coefficients on CaSiO3 and SiO2 using Molecular Dynamics and Experiments} 

\author[1]{D.~Bayer-Buhr\corref{cor1}}
\ead{bayerd@mailserver.tu-freiberg.de}
\ead[url]{https://tu-freiberg.de/fakult4/iwtt/ttd}

\author[2]{M.~Vimal}

\author[2]{A.~Prakash\corref{cor1}}
\ead{Arun.Prakash@imfd.tu-freiberg.de}
\author[1]{U.~Gross}
\author[1]{T.~Fieback}

\address[1]{Technische Universität Bergakademie Freiberg, Chair of Technical Thermodynamics, \\Gustav-Zeuner-Str. 7, 09599 Freiberg, Germany}
\address[2]{Technische Universität Bergakademie Freiberg, Micromechanical Materials Modelling (MiMM), \\Lampadiusstr. 4, 09599 Freiberg, Germany}

\cortext[cor1]{Corresponding author}



\begin{abstract}
{The thermal accommodation coefficient $\alpha$ has been assumed, although lacking any experimental proof, to be near unity for most gases so far, which denotes no influence. However, it plays a contributing role in the field of the effective thermal conductivity of highly porous insulation materials based on $SiO_2$ or $CaSiO_3$ as it is shown in this work.}
{Besides, this work investigates} a possible influence on $\alpha$ for $Ar, N_2, He$ within parameters like temperature, roughness and contamination as this has not been examined on such materials so far. {More importantly}, it answers the question whether the assumption of $\alpha$ = 1 is {valid}. \\
By using a parallel plates device, very similar to the guarded-hot-plate, following EN 12667 it was possible to determine $\alpha$ on a dense $CaSiO_3$. It {occured} that the assumptions $\alpha$ = 1 (for $Ar, N_2$) and $\alpha$ = 0.3 (for $He$) are valid for measurements near room temperature. Further, physical adsorption was found to increase $\alpha$. The determination of the influence of roughness has been started showing an interesting effect, but it still remains an open topic.\\
In a collaborative study molecular dynamics (MD) simulations were performed showing a strong equivalence of $\alpha$ between $SiO_2$ and $CaSiO_3$. {These} results can be considered a lower limit of $\alpha$ as neither roughness nor adsorption processes have been included in the simulation. Therefore, any deviations between experiments and MD could be considered as an apperance of physical adsorption.
\end{abstract}


\end{frontmatter}


\section{Introduction}

Facing today's need for increased energy efficiencies of industrial processes, the thermophysical properties of insulating materials have evoked great interest for decade{s}. Especially new kinds of insulation materials like aerogels or highly porous calcium silicates {($CaSiO_3$)} show {promise} in reducing heat transfer due to their high porosity and pore sizes in the micro and meso range \cite{Swimm2009}. Unfortunately, there is still a lack of understanding of heat transfer due to the very heterogeneous structure of {such} materials. Some efforts were {made} by Swimm \cite{SwimmDiss} who developed a model for {the} description of heat transfer in such materials. Following Swimm {\cite{SwimmDiss}} it {can} be shown that with the usual parallel and series connection of the various heat transfer mechanisms (solid and gas conduction, radiation) the effective thermal conductivity was expected to be lower than the experimental results {indicated}. With introduction of a coupling factor dependent on the material, their model {reflects real-world} experimental results. \\
One detail {of the model of Swimm \cite{Swimm2009} or others (e.g. Zhao \cite{Zhao2012}),} which remains unverified until now {and may be questioning the models of the effective thermal conductivity}, is the influence of the thermal accommodation coefficient $\alpha$ within the gas conduction, which is assumed to be unity for porous insulation materials {so far} \cite{Swimm2009,Zhao2012,Litovsky1996}. In literature, merely few results are available for silicium oxide ({$SiO_2$,} amorphous \cite{McFall1980}, quartz \cite{Saxena1989}) and none for calcium silicate {providing only a weak basis for the models of the effective thermal conductivity}. Therefore, the goal of this {work} is to shed light on the situation of the thermal accommodation coefficient on amorphous $SiO_2$ and $CaSiO_3$ as they usually are the main components in conventional and highly porous insulation materials. {More importantly, it provides the necessary basis of whether $\alpha$ = 1 is valid}. Furthermore, the influence of temperature, roughness and contamination is examined. \\
As the experimental determination of the thermal accommodation coefficient is always influenced by contributing factors (e.g. surface roughness, adsorption layers), which leads usually to an increase of the thermal accommodation coefficient, a collaborative study using molecular dynamics (MD) was done for comparison {to get a more detailed look} onto the situation at an atomistic level. This {gives} a first hint, in what range of values $\alpha$ could be expected, as the results of MD can be considered as the lower limit of the thermal accommodation coefficient on amorphous $SiO_2$ and $CaSiO_3$. Using MD, many studies have been {conducted} previously to determine the accommodation coefficient values for different solid-gas molecule combinations {(e.\,g. amorphous silica aerogels \cite{zichunyang2016molecule} or amorphous polystyrene \cite{Feng2021}, crystal \cite{sipkens2018effect}). However, the study of Zichunyang et al \cite{zichunyang2016molecule} neglects the influence of the attractive forces of the solid atoms between neighboring nanoparticle interfaces. As explained above, Swimm \cite{SwimmDiss} was able to clearly show that a coupling factor has to be introduced in order to represent this improved heat transfer. Due to the neighboring nanoparticle interfaces the attractive forces tend to overlap and increase, which leads to a higher adsorption potential and thus to a higher value of $\alpha$. }
Apart from predicting the values of $\alpha$, the influence of surface roughness \cite{sun2011three}, gas-wall interaction strength \cite{mohammad2020influence, goodman1980thermal, sipkens2018effect}, gas molecular mass \cite{goodman1980thermal,liang2014parametric} and gas temperature \cite{liang2014parametric} on $\alpha$ have also been studied using MD. Higher solid wall surface roughness increases the number of gas-wall collisions through interaction of gas molecules with the lateral edges of the solid atoms \cite{sipkens2018effect}, thus increasing $\alpha$ values. Therefore, MD has shown to be a very good, although also very time-consuming tool, for completing the picture of thermal accommodation coefficients on solid surfaces in comparison to experimental observations.\\
The {paper is structured} as follows: {F}irst, an overall description of the examined insulation materials is given as well as a short summary of the setup of the experimental device. After this a brief description of the MD is given. The final comparison and discussion with experimental and MD results will be presented at last.

\section{Characterization of {P}orous {M}edia}
The materials which were examined are conventional highly porous insulation materials for high temperature usage purposes as well as sintered samples {composed of} $SiO_2$ and $CaSiO_3$ (see Table \ref{tab:Material} for composition). The pore size, {that} show{s} major effects on the pressure dependent effective thermal conductivity $\lambda_{eff}$ \cite{Swimm2009}, ranges from macropores for the porous calcium silicates with a sharp pore size distribution to macro- and mesopores for the porous silicium oxides with a broad pore size distribution. During the experiments the Knudsen effect, where $\lambda_{eff}$ begins to develop a pressure-dependency \cite{Raed2007}, appeared by lowering the pressure. Measurements of the surface roughness ($R_z$) revealed no clear correlation between porosity and roughness, although all highly porous samples were treated the same way (trimming). Sample $CaSiO_3$ D1 was polished with a diamond abrasive paper, but didn't show any significant difference in roughness in comparison to the other samples. \\
All samples are 100\,mm in diameter with a height of 5-7\,mm. Four thermocouples boreholes ($\o$ 1.1\,mm) were drilled horizontally to minimize heat losses by leading the thermocouples along isotherms.
\begin{table}[htpb]
    \centering
\caption{Pore sizes, porosities (measured with mercury intrusion porosimetry), composition and surface roughness ($R_z$, measured with tactile incision technique) of used materials}
\label{tab:Material}
 \begin{tabular}{l|llll}\hline
 material   &  pore size [nm]& porosity [\%] & surface roughness [$\mu$m]& composition \\\hline 
 \textit{$CaSiO_3$ A}          &    320\, \& 700 & 88 & 17.705 &  80\% $CaSiO_3$, 10\% $Zr$,\\ 
 &&&& 10\% Residuals \\
 \textit{$CaSiO_3$ B}          &   530 \& 1360    & 88 & 26.423& 91\% $CaSiO_3$, 9\% Residuals\\ 
 \textit{$CaSiO_3$ C}          &    350          & 86 & 39.706&100\% $CaSiO_3$\\  
 \textit{$CaSiO_3$ D1}           &n.a.      &   less than 5  &36.120 &100\% Wollastonite\\
 \textit{$CaSiO_3$ D2}           &n.a.      &   less than 5 & n.a. & 100\% Wollastonite \\
 \textit{$SiO_2$ A}          &   22 \& ca. 2000  & 81 &  24.875 & 80\% $SiO_2$, 15\% $ZrSiO\textsubscript{4}$,  \\ 
 &&& &5\% Residuals \\
 \textit{$SiO_2$ B}          &   15 \& ca. 780  & 86  &19.865& 80\% $SiO_2$, 20\% $SiC$\\\hline
 \end{tabular}
\end{table}

Concerning the preparation of the sintered sample ($CaSiO_3$ D1, wollastonite) the following procedure has been performed: 250\,g wollastonite\footnote{Wollastonite is the natural form of $CaSiO_3$.} Casiflux (Sibelco) and 2.5\,g TiO\textsubscript{2} R 320 (Sachtleben) (used as a sintering aid) were mixed with 100\,g deionized water for at least 1 hour. In order to obtain a castable slurry, additives were added to the slurry: 2.5\,g Optapix AC 170 (Zschimmer \& Schwarz, Germany), 0.5\,g Axilat RH 50 MD (C.H. Erbslöh, Germany), and 1\,g Castament F60 (BASF Construction Solutions GmbH, Trostber, Germany). The Optapix AC 170 was used as a binder, the Castament F60 acted as dispersant and the Axilat were added to change the dilatant behavior of the slurry to obtain a shear thinning behavior which is necessary for the {casting} process. The slurry was {cast} into the casting mold, which is basically of the same form as the porous samples ($\o$ 100\,mm, height 7\,mm). The sample was dried carefully {at} 50\textcelsius{} and sintered at 1200\textcelsius{} for 3 hours. The porosity is assumed to be significantly lower (less 5\%) than for the other porous samples.

\section{Experimental {Setup and Mathematical Foundation}}

For the determination of thermal accommodation coefficient{s} on amorphous $SiO_2$ and $CaSiO_3$ a measurement device similar to {the} well-known steady-state guarded-hot-plate device \cite{EN12667} with parallel plates was developed and built. {It} will be presented in the following. After this a more detailed description of the test procedure and the extraction of the thermal accommodation coefficient from experimental measurements are given.

\subsection{Setup}
\begin{figure}[htpb]
 \centering
 \includegraphics[width=14.5cm]{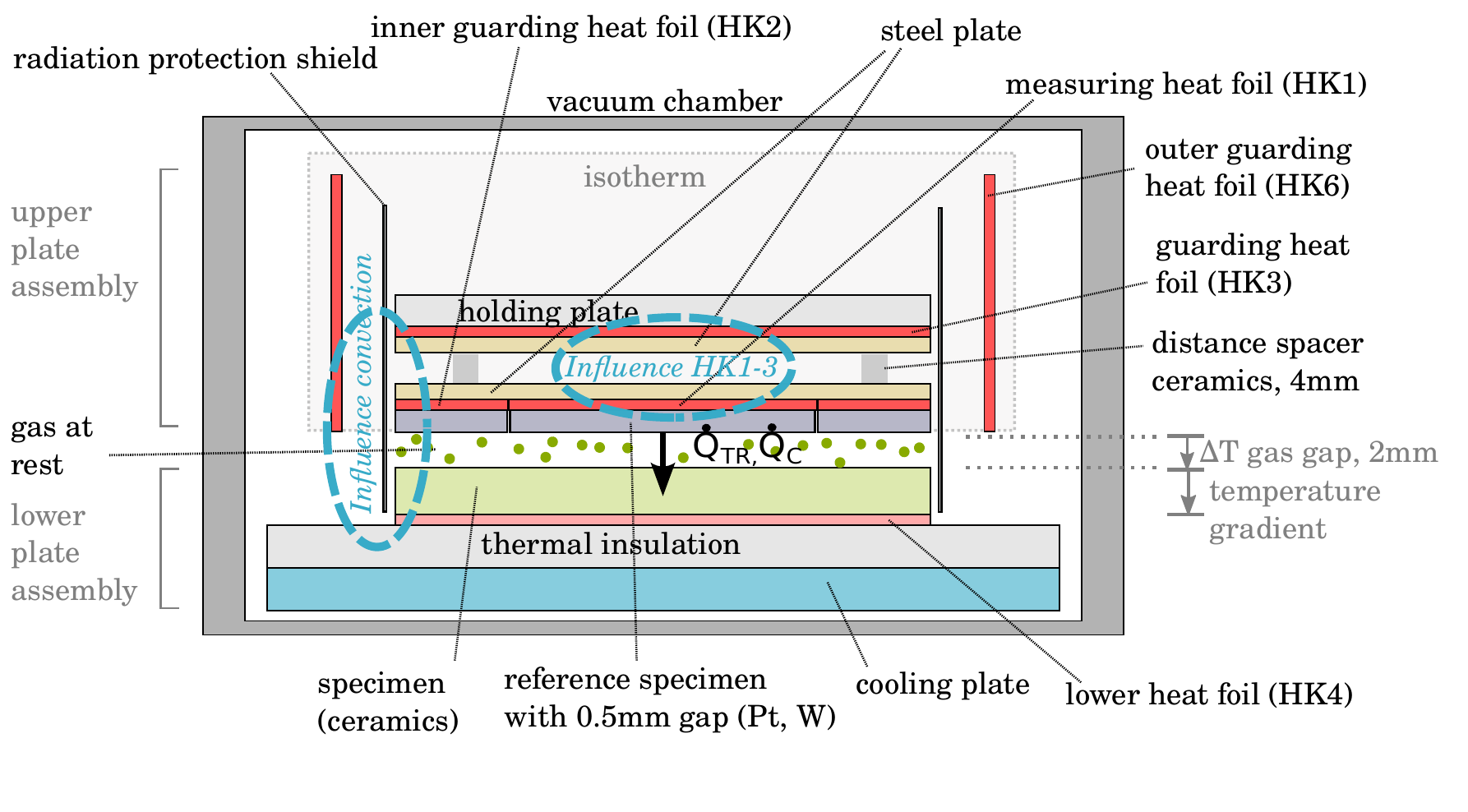}
 \caption{Scheme of the developed measurement device, similar to guarded-hot-plate regarding \cite{EN12667}; $\dot{Q}_{TR}$ is the heat flow in the transition regime, $\dot{Q}_C$ is the heat flow in the continuum regime; each measurement is performed at least twice to ensure repeatability; the temperature level is t\textsubscript{upper}/t\textsubscript{lower} = 75/65$^\circ$C}
 \label{fig:Schema_Anlage}
\end{figure}

A schematic of the experimental setup is shown in Figure \ref{fig:Schema_Anlage}. The underlying idea of this method is the use of guarding heat foils (HK2/3/6) based on superthin polyimide foil (less than 0.2\,mm thick) surrounding the measuring heat foil (HK1) in the center. Thus, an isothermal zone is created to ensure that the heat flow which is needed for the derivation of the thermal accommodation coefficient is one-dimensional through the gas gap (2\,mm). The temperature difference over the gas gap is 10\,K controlled by the underlying heat foil (HK4). The gas gap is confined above by a reference sample which is held {in place} by the upper plate assembly and below by the ceramic sample which is placed above HK4. Therefore, the thermal accommodation coefficient of the ceramic material $\alpha_c$ is determined in dependence of the thermal accommodation coefficient of the reference sample $\alpha_r$. The latter is determined during calibration. For thermal separation the upper plate assembly is held in position by a linear translator which is placed above the vacuum chamber. The vacuum chamber is cylindrical with a height of 300\,mm and a diameter of 200\,mm. It can be opened vertically and it holds a pressure of 10\textsuperscript{-4}\,mbar. The essential water-cooled heat sink is placed below heat foil HK4 and provides the lowest temperature in the system (for more details see \cite{Bayer2014}).


\subsection{Test {P}rocedure and {D}etermination of $\alpha$}

\begin{figure}[htpb]
\centering
\includegraphics[width=11cm]{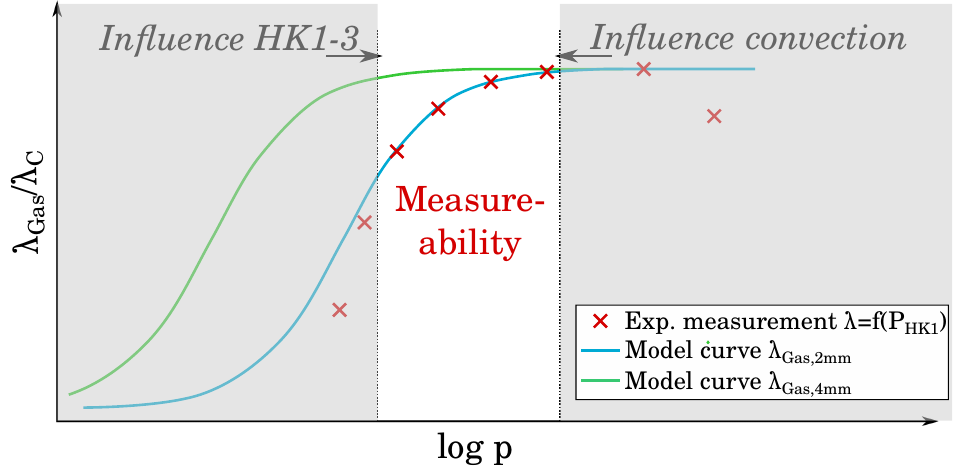}
\caption{Schematic measureability of the measuring device; two occuring phenomenon leading to upper and lower limit of measurability; y-axis: $\lambda_{Gas}$ is the gas thermal conductivity at arbitrary pressure, $\lambda_C$ is the gas thermal conductivity in continuum range, x-axis: logarithmic pressure $p$; $P_{HK1}$ = power of HK1 }
\label{fig:Messfahigkeit1}
\end{figure}

The measurements start with baking-out at 140$^\circ$C and evacuation down to the lowest possible pressure (ca.~10\textsuperscript{-4}\,mbar) for at least 12\,h. Further the atmosphere is flushed and evacuated {twice more} very rapidly to make sure that gas residuals are flushed out of the vacuum chamber. After this atmospheric cleaning process, four steady-state conditions at different pressures are generally reached at a constant temperature of at least 60\,minutes, each between 3 - 30\,mbar for \textit{Ar} and $N_2$ as well as 5 - 300\,mbar for $He$. During the measurement all signals like current and voltage for calculation of the power of HK1 as the heat flow rate applied for the thermal accommodation coefficient determination, temperature readings of about 30 thermocouples, and inside pressures are logged every 10\,s. The measured pressure value is needed to control the stability of the gas regime and for {the} calculation of the $Kn$-number. Additionally the thermocouples provide an overall picture of the thermal homogeneity to ensure the validity of the underlying parallel plate principle. {In order to separate the radiation component from the measured heat flow, the radiation is estimated and subtracted from the measured heat flow. A measurement at very low pressures without any influences of gas thermal conduction was not possible, as will be explained below. (Rev 2, pt 4)}\\
Two phenomena which were observed during calibration led to a confined measurability and will be presented more in detail in the following as this helps {to get} an impression on the challenging aspects of this project:  
\paragraph{Mutual influences between HK1 and HK3}  The task of HK3 is to provide a $\Delta T$ = 0\,K (temperature difference) next to HK1 as the measuring heat foil. Unfortunately, it turned out that the thermal resistivity of the layer in between HK1-3 had to exhibit a very high level to fulfill this task. After testing different materials the best option was to create a larger gas gap of 4\,mm (see Figure \ref{fig:Schema_Anlage}) by using ceramic distance spacers. The idea for this can be followed in Figure \ref{fig:Messfahigkeit1}, where the well-known S-curve of the pressure dependent gas thermal conductivity can be seen versus logarithmic pressure: The gas thermal conductivity in the 2\,mm gap varies during the measurement depending on the pressure (blue curve) whereas in the 4\,mm gap (green curve) there is an almost pressure-independent gas thermal conductivity. Thus, only the 2\,mm gap gas conduction shows a pressure-dependent behaviour which is needed to derive the thermal accommodation coefficient. Below the lower pressure limit an influence of the pressure-dependent thermal conductivity within the 4\,mm gap starts to develop, which leads to an incorrect measurement of the heat flow of HK1, which is needed for the derivation of $\alpha$. This phenomenon therefore defines a lower pressure limit for the measurability of the device. 

\paragraph{Influence of convection} Another phenomenon defines the upper limit of measurability of the device. The radiation protection shield consists of a very thin metal foil, which is very flexible. Therefore, due to all the wiring of thermocouples, it is not possible to place the shield in an exact position. By means of numerical investigations it can be shown that near atmospheric pressure natural convection develops in the annulus. Since the gap between the radiation protection shield and the thermal insulation below HK4 is not completely closed, an accelerated flow occurs at the outer edge of HK2, which leads to increased heat losses. Since HK2 has to compensate for this heat loss in order to maintain $\Delta T$ = 0\,K, the power of HK2 increases. At the same time, the power of HK1 decreases because it is not exposed to the same heat loss as HK2. Due to this, the power of HK1 decreases at higher pressures, as can be seen in Figure \ref{fig:Messfahigkeit1}. As the lower and upper limit are also dependent on the kind of gas, different upper pressure limits could be found ranging from 3\,mbar for $Ar$ up to 300\,mbar for $He$.

For calculation of the thermal accommodation coefficient the model of Springer \cite{Springer1971} is followed, which extends the closed-form solution of heat conduction in rarefied gases between parallel plates developed by Lees and Liu \cite{LeesLiu1960} to arbitrary values of the thermal accommodation coefficient:
\begin{equation}
 \frac{\dot{Q}_{TR}}{\dot{Q}_{FM}} = \Big(1+\frac{4}{15}\frac{1}{Kn} \frac{\alpha_r \alpha_{cer}}{\alpha_r+\alpha_{cer} - \alpha_r \alpha_{cer}}\Big)^{-1}
\end{equation}

\noindent with $\dot{Q}_{TR}$ and $\dot{Q}_{FM}$ as the heat flow rates by conduction through the gas in the transition regime and in the free molecular regime. {R}espectively, $Kn$ {is} the Knudsen number and $\alpha_r$ and $\alpha_{cer}$ {is} the thermal accommodation coefficients on the upper/reference and lower/ceramic plate. With the help of Sherman's interpolation \cite{Sherman1963},
\begin{equation}
 \frac{\dot{Q}_{TR}}{\dot{Q}_{FM}} = \Big(1+ \frac{\dot{Q}_{FM}}{\dot{Q}_C}\Big)^{-1}
\end{equation}

where $\dot{Q}_{C}$ is the heat flow rate by conduction through the gas in the continuum regime, one can obtain the following equation for experimental determination of the thermal accommodation coefficient (at constant temperature difference):
\begin{equation}
 \frac{\dot{Q}_{TR}}{\dot{Q}_{C}} = \Big(1+\frac{15}{4} Kn \frac{\alpha_r+\alpha_{cer} - \alpha_r \alpha_{cer}}{\alpha_r \alpha_{cer}}\Big)^{-1}
 \label{eq:Springer1}
\end{equation}

As $\alpha$ is always influenced by the temperatures of the gas and wall molecules as well as by the composition, the surface structure and eventual contamination of the wall material, two different accommodation coefficients are present in a parallel plates setup. If the surfaces confining the gas gap {are} similar (same material and surface treatment) one can asume: $\alpha_{r,1} = \alpha_{r,2} = \alpha_r$. This gives a simpler version of Equation~\ref{eq:Springer1}:
\begin{equation}
 \frac{\dot{Q}_{TR}}{\dot{Q}_{C}} = \Big(1+\frac{15}{4} Kn \frac{\alpha_{r,1}+\alpha_{r,2} - \alpha_{r,1} \alpha_{r,2}}{\alpha_{r,1} \alpha_{r,2}}\Big)^{-1} = \Big(1+\frac{15}{4} Kn \frac{2-\alpha_r}{\alpha_r}\Big)^{-1}
 \label{eq:Springer2}
\end{equation}

As observed during measurements the temperature difference across the gas gap changes due to a pressure-dependent contact resistance between the heat foil and the sample for the lower plate. Therefore, this influence has to be corrected leading to the following equation, which is used for calibration:
\begin{equation}
 \boxed{\frac{\dot{Q}_{TR}}{\dot{Q}_{C}}*\frac{\Delta T_C}{\Delta T_{TR}} = \Big(1+\frac{15}{4} Kn \frac{\alpha_{r,1}+\alpha_{r,2} - \alpha_{r,1} \alpha_{r,2}}{\alpha_{r,1} \alpha_{r,2}}\Big)^{-1} = \Big(1+\frac{15}{4} Kn \frac{2-\alpha_r}{\alpha_r}\Big)^{-1}}
 \label{eq:Springer3}
\end{equation}

where $\Delta T_{C}$ and $\Delta T_{TR}$ represent the temperature differences over the gas gap in the continuum and transition regime, respectively. By measuring the ceramic sample the equivalent equation is used:
\begin{equation}
 \boxed{\frac{\dot{Q}_{TR}}{\dot{Q}_{C}}*\frac{\Delta T_C}{\Delta T_{TR}} = \Big(1+\frac{15}{4} Kn \frac{\alpha_{r}+\alpha_{cer} - \alpha_{r} \alpha_{cer}}{\alpha_{r} \alpha_{cer}}\Big)^{-1} }
 \label{eq:Springer4}
\end{equation}

By choosing a certain thickness of the gas gap (here: 2\,mm) and different pressure values the transition and continuum gas state can sequentially be reached in the vacuum chamber. For finally calculating $\alpha_{r}$ or $\alpha_{cer}$ they are determined equivalently based on Equation \ref{eq:Springer3} or \ref{eq:Springer4}, respectively, where $\alpha_{r}$ or $\alpha_{cer}$ is fitted onto the experimental results with the help of the nonlinear Levenberg-Marquardt regression method \cite{octave-leasqr}.

%
%
%
%


\section{Molecular Dynamics Study}

A setup similar to the experimental one is created and presented below together with a description of the method used to calculate $\alpha_{cer}$. Finally, used potentials and parameters, that have been validated with literature are shown. To validate the used calculation method of $\alpha_{cer}$ on amorphous $SiO_2$ and $CaSiO_3$ the method was tested with {tungsten} ($W-Ar$ and $W-He$), since reliable literature data are available for these.

\subsection{Setup}
\label{sec:MD-Setup}
\begin{figure}[htpb]
    \centering
    \includegraphics[width=15cm]{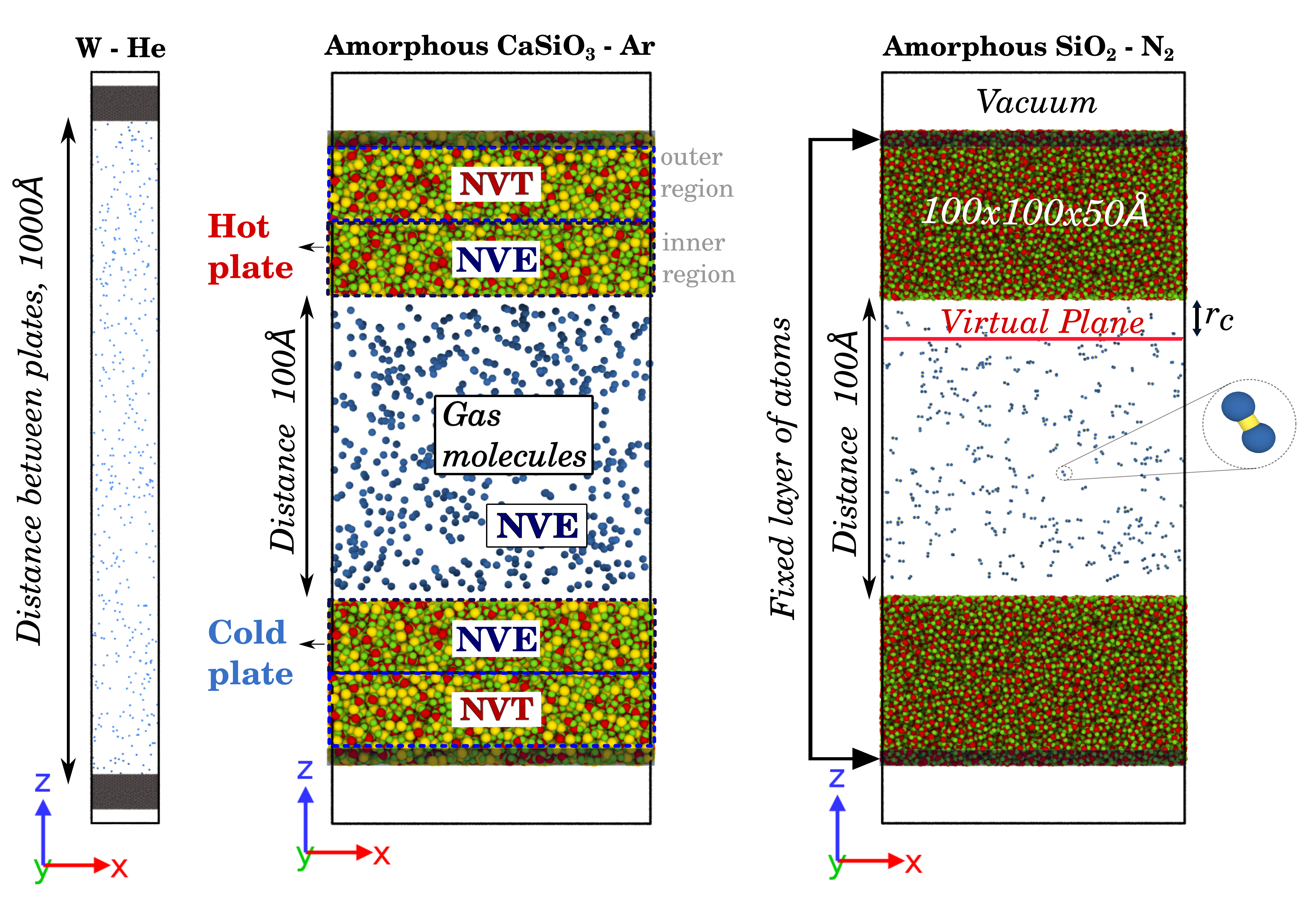}
    \caption{Representative overview of MD cases for amorphous $SiO_2$, $CaSiO_3$ and crystalline $W$; parallel plates setup; NVE/NVT for all cases; position of virtual plane at one cutoff-distance $r_c$ for all cases; simulation for all $He$-cases with wider distance to ensure similar $Kn$-numbers in comparison to $Ar/N_{2}$-cases (for $He$ - 1000\,\si{\angstrom}, for $Ar$ and $N_{2}$ - 100\,\si{\angstrom}); Other cases ($CaSiO_3$-$He$, $CaSiO_3$-$N_{2}$, $SiO_2$-$Ar$, $SiO_2$-$He$, $W-Ar$) follows identical simulation setups as in the representative cases; Color code for atoms: red-silicon, green-oxygen, yellow-calcium, black-tungsten.}
    \label{fig:MD-Setup}
\end{figure}

\begin{figure}[htpb]
\centering
\fontsize{9}{11}\selectfont 
 \input{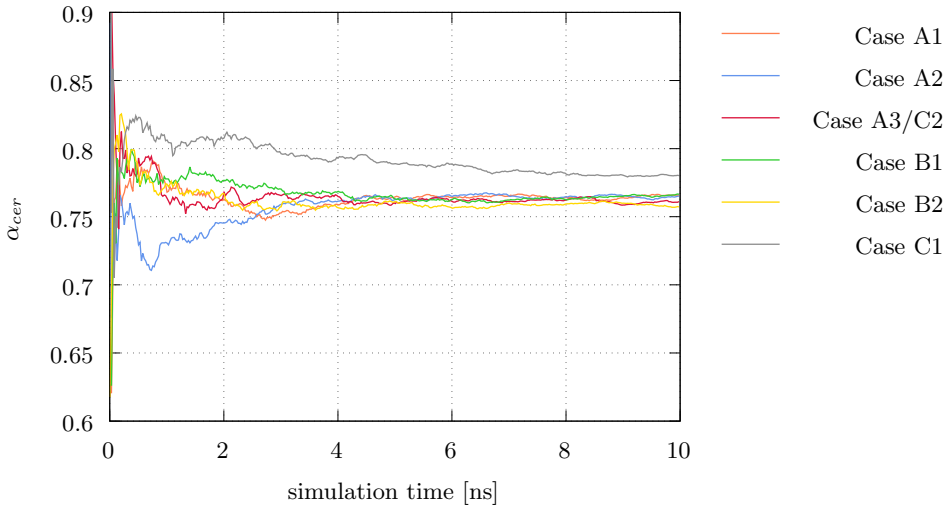}
\caption{Influence of thermostats, inital seed velocity distribution and temperature level on $\alpha_{cer}$; Case A1: Berendsen thermostat, $\alpha_{cer}$ = 0.765; Case A2: Langevin thermostat - $\alpha_{cer}$ = 0.766; Case A3/C2: Nos\'e-Hoover thermostat/temperature level 340/350\,K - $\alpha_{cer}$ = 0.761; Case B1: initial seed velocity distribution 1 - $\alpha_{cer}$ = 0.766; Case B2: initial seed velocity distribution 2 - $\alpha_{cer}$ = 0.757; Case C1: temperature level 310/320\,K - $\alpha_{cer}$ = 0.78}
\label{fig:thermostats}
\end{figure}

For generating the amorphous structures of $SiO_2$ and $CaSiO_3$, the melt-quench process is followed. The starting structure is created by randomly distributing 37,500 ($CaSiO_3 \rightarrow$ 7,500 - \textit{Ca}; 7,500 - \textit{Si}; 22,500 - \textit{O}) or 38,400 ($SiO_2 \rightarrow$ 12,800 - \textit{Si}; 25,600 - \textit{O}) ions in a box of $\approx$ 10 x 10 x 5 nm based on the reference densities \cite{mead2006molecular,munetoh2007interatomic}  with periodic boundary conditions. The time-temperature cycle for the entire melt quench process considered by Mead \cite{mead2006molecular} ($CaSiO_3$) or Munetoh \cite{munetoh2007interatomic} ($SiO_2$) is followed. Tungsten, with a body centered cubic structure and same sample dimensions as in the amorphous structure case, is constructed by orienting the densely packed plane (110) along the z direction of the simulation box. The structures are well equilibrated at 300\,K and at zero pressure in a NPT ensemble for 160\,ps. These final equilibrated structures of amorphous $SiO_2$, $CaSiO_3$ and $W$ are used as the solid substrates for their respective simulations (see Figure \ref{fig:MD-Setup}). \\
The distance between the upper (hot) and lower (cold) plate is maintained at 100\,\si{\angstrom} for simulations involving $Ar$ \& $N_{2}$ gas molecules and 1000\,\si{\angstrom} for $He$ gas molecules to maintain the total gas pressure below their corresponding critical pressures. The solid plates are placed in such a way that the gas interacting surfaces have identical amorphous surfaces at the top and bottom, i.e. {the} upper plate is just the mirror copy of the lower plate about the center of simulation box. Periodic boundary conditions are applied along x, y and z directions to facilitate the use of PPPM long range solvers. The atoms present within 3.0\,\si{\angstrom} distance from the outermost layer of walls are fixed to prevent the walls from sliding along the z direction due to gas pressures. Along {the} z direction for a distance of 20\,\si{\angstrom} beyond the fixed layers of atoms, vacuum is maintained to prevent energy exchange between the hot and cold plates across the periodic boundary. \\
In an equilibrium MD for computing $\alpha_{cer}$, the temperatures of the hot and cold plates {need} to be maintained at a preset temperature throughout the complete production run. Thermosetting the entire solid plates using a deterministic Nos\'e-Hoover (NH) thermostat \cite{nose1984unified, hoover1985canonical}, which can preserve the structural properties, could be a viable option. But since the energy/momentum exchange between the solid \& gas molecules is the predominant process in this simulation, the NH thermostat preserves the dynamical properties only under certain circumstances. It is indeed a recommended practice to turn off the thermostats before the production run for applications where the dynamical properties are sampled \cite{braun2019best}. So, in order to maintain the temperature and to have the gas-wall interaction region to be more physical by preserving both the structural as well as the dynamical properties during the production run, the solid plates\, are\, split \,equally 
\newpage
into  two regions along the z direction where one region is maintained under micro-canonical (NVE) ensemble that contains the gas-wall interaction surface and the other maintained under the canonical (NVT) ensemble conditions where the temperatures are controlled using Nos\'e-Hoover thermostat throughout the entire simulation time. An initial equilibration time of 500\,ps is allowed for the heat to transfer from the outer to inner region, such that both the NVT \& NVE regions {attain} the same temperature. \\
The outer regions of the hot and cold plates are maintained at 350\,K and 340\,K, respectively, to reflect experimental conditions. Further, the influence of temperature levels close by (+/-20\,K) did not reveal a great effect on $\alpha_{cer}$ and so this parameter variation is neglected in the following (see Figure \ref{fig:thermostats}).\\
After the equilibration of solid plates, the gas molecules are introduced in-between the plates at random positions. An appropriate gas molecule density has to be chosen due to the trade-off involved between increasing the number of gas-wall collisions within the finite simulation time and maintaining the gas pressure below the critical pressure. The number of gas molecules to be introduced is determined by considering the critical pressure of the gases (p\textsubscript{cr} = 4.86\,MPa for $Ar$, p\textsubscript{cr} = 0.23\,MPa for $He$, p\textsubscript{cr} = 3.4\,MPa for $N_{2}$) as the upper limit and also by taking the Knudsen number (Kn = 0.3 for $Ar$ and $N_{2}$ and Kn = 0.65 for $He$) in the transition flow regime into account.\\ 
During both the solid and gas molecules initialization, in order to obtain the Maxwell-Boltzmann distribution for the velocities, random values are sampled from a Gaussian distribution for each velocity component with a mean of 0.0 and a standard deviation scaled to their respective temperatures.\\
The gas molecules are maintained under NVE ensemble conditions and 1\,ns of simulation time is allowed to equilibrate the gas temperature. The extraction of the trajectories of gas atoms for computing $\alpha_{cer}$ starts only after the 1.5\,ns of the total equilibration time. 

\subsection{Calculation of $\alpha_{cer}$}

For computing of $\alpha_{cer}$, a virtual plane (see Figure \ref{fig:MD-Setup}) is constructed at a distance of one gas-wall interaction cut off radius (r\textsubscript{c}) away from the inner surface of the upper solid wall normal to the z direction, to avoid any influence of the attractive forces from the solid wall atoms \cite{sun2011three,mohammad2020influence,liang2014parametric}. Gas atoms crossing the virtual plane towards the solid wall {with} positive z velocities (v$_z$ $>$ 0) are categorized as incident atoms, which represents the start of the energy exchange process. {The} same gas atoms recrossing the virtual plane after having {collided} with the solid wall {with} negative z velocities (v$_z$ $<$ 0) are categorized as reflected atoms, which represents the end of the energy exchange process. This is based on the assumption that the probability of gas-gas collisions within the upper solid wall and the virtual plane region is negligible, i.e. gas atoms crossing the virtual plane will return back only after having direct collision with the solid wall atoms \cite{liang2014parametric}. Only the velocities of gas molecules crossing this virtual plane at any time step are considered for calculating $\alpha_{cer}$. For the case of monoatomic molecules such as $Ar$ and $He$, the positional coordinates and velocities of atoms crossing the virtual plane are directly exported whereas for the case of diatomic molecule{s} such as $N_{2}$, the  positional coordinates and velocities at the geometric center of mass (COM) of the diatomic molecules \cite{zichunyang2016molecule} are exported and only based on the COM, a molecule is checked whether it has crossed the virtual plane or not during the simulation. The formula for calculating $\alpha_{cer}$ from Spijker \cite{spijker2010computation}, based on the least squares approximation on the collision data containing incident ($K_I$) and reflected ($K_R$) quantities is given by,
\begin{equation}
\alpha_{cer} \ = \ 1 \ - \ \frac{\sum_{i} \ (K_I^i-\left\langle K_I \right\rangle)\ (K_R^i-\left\langle K_R \right\rangle)}{\sum_i \ (K_I^i-\left\langle K_I \right\rangle)^2} 
\end{equation}

where the summation is performed over all the recorded collisions for the considered quantity (kinetic energy). To allow a sufficient number of gas-wall collisions so as to achieve good statistics for the $\alpha_{cer}$ calculations, the total simulation time is chosen based on the convergence of {the} $\alpha_{cer}$ value with respect to simulation time. All the simulations are performed with a time step of 1\,fs (except 0.5\,fs for the amorphous $SiO_2$ system) using the LAMMPS package \cite{plimpton1995fast} and post processing \& visualization have been done using the open source visualization tool OVITO \cite{stukowski2009visualization}.

\subsection{Interatomic Potentials and {P}arameters}

The interatomic potentials used for the amorphous $SiO_2$, $CaSiO_3$ and $W$ as well as for the gases $Ar$, $He$ and $N_{2}$ are listed in Table \ref{tab:Potentiale}. The potential expression of $CaSiO_3$ \cite{pedone2006new} involves a dispersion term in addition to a long{-}range Coulombic term and a short-range Morse function. The addition of the repulsive term is necessary to model interactions at high temperatures and pressures. Simulating using potentials without the repulsive contribution term for cases where the ions tend to come close enough, i.e. at elevated temperatures and during interactions with fixed layer of atoms, could
be problematic due to the unphysical attraction values at low inter-ionic distances. For the evaluation of long range Coulombic interactions, {a} particle-particle-particle mesh (PPPM) over the standard Ewald summation method is chosen for its computational efficiency with a force norm of less than 10\textsuperscript{-6}\,eV/\si{\angstrom}. To avoid the computation of long range coulombic interactions, {a} 3-body Tersoff potential for $SiO_2$ \cite{munetoh2007interatomic} over the Beest-Kramer-Santen (BKS) or Morse potential is used. Lennard-Jones (LJ) [12-6] potential \cite{jones1924determination} is widely used to model gas-gas and gas-wall interactions. Each $N_{2}$ molecule is modeled as a rigid rotor with a fixed bond length of 1.1\,\si{\angstrom} \cite{liang2014parametric}. The parameter values for gas-wall interactions can be calculated by using the approximation methods like Fender-Halsey (FH) and Lorentz-Berthelot (LB) mixing rules. Since {the} LB mixing rule is observed to overestimate the gas-wall atoms bonding characteristics \cite{mohammad2020influence}, parameter values in the current simulations are calculated using {the} FH mixing rule. Cut-off radius for gas-gas and gas-wall interactions are 2.5\,times $\sigma_{ii}$ \cite{mohammad2020influence} and 8\,\si{\angstrom}, respectively. The LJ parameter values chosen for $Ca$, $Si$ and $O$ atoms are successfully used to study the physical adsorption behavior of gases on to the amorphous silica, silicate glass surfaces and zeolites \cite{bakaev1999adsorption, talu2001reference}. LJ parameter values ($\varepsilon$\textsubscript{W-Ar} = 0.002168\,eV \& $\sigma$\textsubscript{W-Ar} = 2.93\,\si{\angstrom}) used in the tungsten surface scattering studies \cite{ozhgibesov2013studies} are employed for {$W-Ar$} interactions. For $W-He$, the functional form and parameter values used in tungsten and helium interaction studies \cite{juslin2013interatomic} {are} chosen. The cutoff {radii} for tungsten-argon and tungsten-helium interactions are 6\,\si{\angstrom} and 4\,\si{\angstrom} respectively. All potentials and parameter values are substantially tested and validated with their corresponding literature. 

\begin{table}[htbp]
\caption{Overview of interatomic potentials and parameters used for MD; all validated with literature}
 \label{tab:Potentiale}
    \begin{tabular}{P{2.5cm}|P{2.5cm}P{2.5cm}P{2.5cm}P{2.5cm}}  
    \hline
    \multicolumn{5}{c}{\textbf{Interatomic potentials}}  \\\hline
    $CaSiO_3$ \cite{pedone2006new}& \multicolumn{4}{P{12.5cm}}{$U(r) \ = \ 	\frac{z_iz_je_{ecc}^2}{r} \ + \ D_{ij}\left\lbrace[1-e^{-a_{ij}(r-r_0)}]^2-1\right\rbrace \ + \ \frac{C_{ij}}{r^{12}}$


where \emph{D\textsubscript{ij}} is the bond dissociation energy, \emph{a\textsubscript{ij}} is a function of the slope of the potential energy well, \emph{C\textsubscript{ij}} is the repulsive contribution parameter, \emph{r\textsubscript{0}} is the equilibrium bond distance, \emph{z} is the ionic charge and \emph{e\textsubscript{ecc}} is the elementary charge constant. }\\

Ion pair & D\textsubscript{ij} [eV]& a\textsubscript{ij} [\si{\angstrom}\textsuperscript{-1}]& r\textsubscript{0} [\si{\angstrom}]& C\textsubscript{ij} [eV \si{\angstrom}\textsuperscript{12}]\\\hline
			$Si^{+2.4} - O^{-1.2}$ & 0.340554 & 2.006700 & 2.100000 & 1.0 \\
			$Ca^{+1.2} - O^{-1.2}$ & 0.030211 & 2.241334 & 2.923245 & 5.0 \\
			$O^{-1.2} - O^{-1.2}$ & 0.042395 & 1.379316 & 3.618701 & 22.0 \\
			\hline

$SiO_2$ \cite{munetoh2007interatomic} &\multicolumn{4}{P{12.5cm}}{$E \ = \ \frac{1}{2} \ \sum_i \ \sum_{j\neq i} \ V_{ij}$ 

where $E$ is the potential energy (more details in \cite{tersoff1989modeling}).}\\
& \multicolumn{4}{c}{Parameter set developed by Munetoh \cite{munetoh2007interatomic} is used.}\\\hline

$W$    \cite{daw1984embedded}    &\multicolumn{4}{P{12.5cm}}{$E \ = \ \sum_{i} \ G_i \> \left(\sum_{j\ne i}\rho_j^a(r_{ij})\right) \ + \ \frac{1}{2} \ \sum_{i,j (j\ne i)} U_{ij}(r_{ij}) $ 

where $r_{ij}$ is the distance between atoms \emph{i} \& \emph{j}, \emph{G} is the embedding energy, $\rho^a$ is the spherically averaged atomic electron density and \emph{U\textsubscript{ij}} is the pair-wise potential interaction between atoms \emph{i} \& \emph{j}. For defining the interaction between tungsten atoms, the EAM potential parameters developed by Olsson \cite{olsson2009semi} {have} been used.} \\\hline

\multicolumn{5}{c}{\textbf{Gas-gas-interactions}} \\\hline

$Ar, He, N_2$ \cite{jones1924determination}  & \multicolumn{4}{P{12.5cm}}{$U(r\textsubscript{ij}) \ = \ 4\varepsilon \Big[ \Big( \frac{\sigma}{r\textsubscript{ij}}\Big)^{12} \ - \ \Big( \frac{\sigma}{r\textsubscript{ij}}\Big)^{6} \Big]$ (Lennard-Jones-Potential)

where \emph{$\varepsilon$} is the depth of the potential well, $\sigma$ is the distance at which the interatomic potential between two particles is zero and \emph{r\textsubscript{ij}} is the interatomic distance between the atoms \emph{i} \& \emph{j}.} \\

& \multicolumn{2}{c}{$\varepsilon\textsubscript{ii}$ [eV]} & \multicolumn{2}{c}{$\sigma\textsubscript{ii}$ [\si{\angstrom}]} \\
			\hline
			$Ar - Ar$ \cite{hirschfelder1964molecular}& \multicolumn{2}{c}{0.01032} & \multicolumn{2}{c}{3.4}  \\
			$He - He$ \cite{hirschfelder1964molecular} & \multicolumn{2}{c}{0.00094} & \multicolumn{2}{c}{2.64}   \\
			$N_2 - N_2$ \cite{zambrano2014molecular}& \multicolumn{2}{c}{0.00313} & \multicolumn{2}{c}{3.32}  \\
			\hline
\multicolumn{5}{c}{\textbf{Gas-wall-interactions (for FH-mixing rule with parameters from gas-gas-interactions)}} \\\hline

   All gas/wall materials\cite{Fender1962}     & \multicolumn{4}{P{12.5cm}}{$\sigma_{ij} \ = \ \frac{\sigma_{ii} \ + \ \sigma_{jj}}{2} \ ; \ \ \varepsilon_{ij} \ = \ \frac{2 \varepsilon_{ii}.\varepsilon_{jj}}{\varepsilon_{ii} \ + \ \varepsilon_{jj}} $ 
   
   Fender-Halsey-Mixing-Rule }\\
Ion pair & \multicolumn{2}{c}{$\varepsilon$\textsubscript{ii} [eV]} & \multicolumn{2}{c}{$\sigma$\textsubscript{ii} [\si{\angstrom}]}\\
			\hline
			$Si - Si$ \cite{watanabe1995investigation}& \multicolumn{2}{c}{0.00160} & \multicolumn{2}{c}{0.677} \\
			$Ca - Ca$ \cite{watanabe1995investigation}& \multicolumn{2}{c}{0.01179} & \multicolumn{2}{c}{1.764} \\
			$O - O$ \cite{watanabe1995investigation}& \multicolumn{2}{c}{0.00875} & \multicolumn{2}{c}{2.708} \\
			\hline
    \end{tabular}
 \end{table}

\section{Results}
In the following, the results obtained from MD and experiments for {validation} with W as well as for amorphous $SiO_2$ and $CaSiO_3$ are presented.

\subsection{Validation with {Tungsten - Experiments and Molecular Dynamics}}
\begin{figure}[htpb]
   \centering
      \subfloat[W-Ar]{\fontsize{9}{11}\selectfont\input{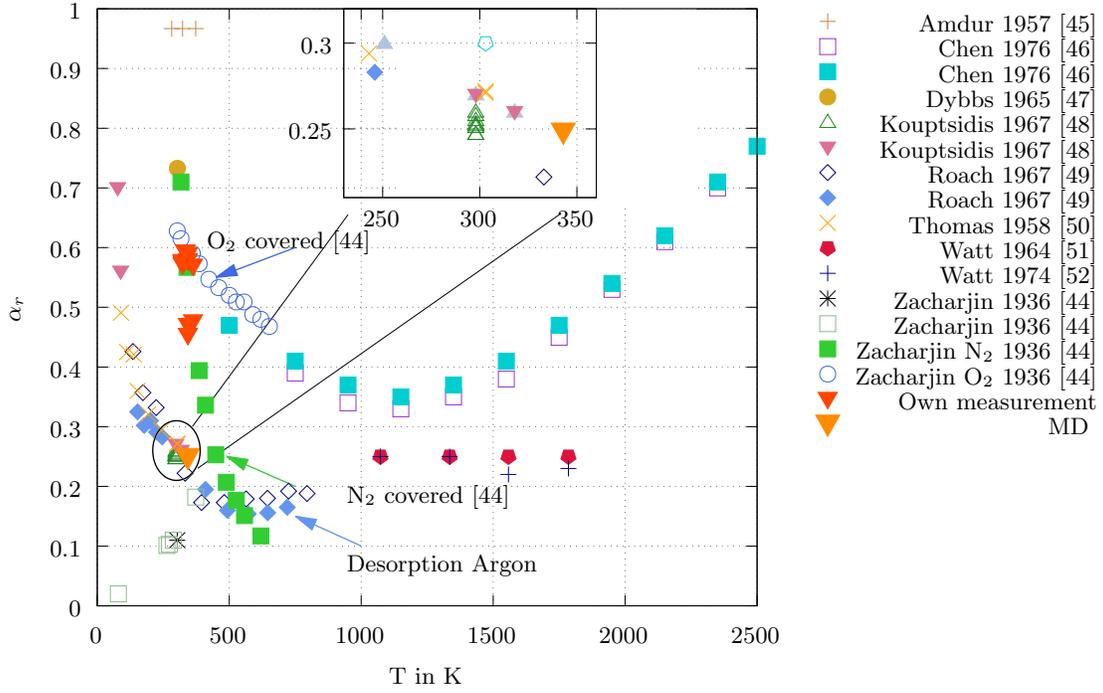}}\qquad
      \subfloat[W-He]{\fontsize{9}{11}\selectfont\input{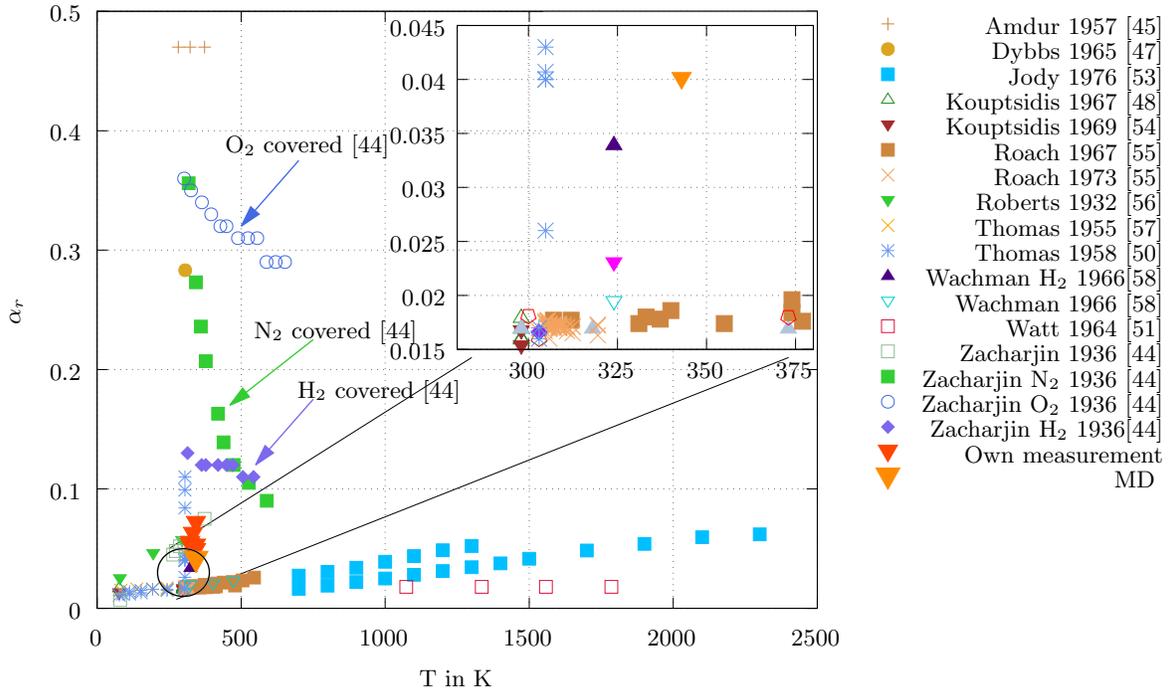}}\qquad
      \caption[Titel des Bildes]{Overview of literature results of $\alpha_r$ over temperature for $W-Ar$ and $W-He$ in comparison with simulation (orange triangles) and experimental results (red triangles, $W$ showed influences of coverage of unknown nature, assumed chemical adsorption): $N_{2}$ coverage and desorption examined by Zacharjin \cite{Zacharjin1936}}
      \label{fig:W-Ar}
\end{figure}

In Figure \ref{fig:W-Ar} literature data in comparison with the results from experiments and MD are shown. Basically the data from literature scatters very much, but some trends are visible. For $W-Ar$, the desorption of $Ar$ is evident with a strong decreasing trend with increasing  temperature above 87\,K, the condensation temperature of $Ar$ \cite{Quantachrome}. In contrast, it is evident that there is no decreasing trend for $W-He$ due to the non-adsorptive behavior of $He$. Furthermore, {literature data varies} between 0 to 1 over the entire temperature range. Usually this behavior is explained with adsorption effects of unknown nature (e.g. Saxena \& Joshi \cite{Saxena1989}). For {the} validation with $W$ (red triangles for exp., orange triangles for MD) the results from MD agree very well with literature data. {T}herefore, the method to derive the thermal accommodation coefficient from MD {(following Spijker \cite{spijker2010computation})} is validated. \\
Further, results from the experimental calibration indicate at least chemical adsorption at the $W$ surface leading to a higher $\alpha_r$. Steinbeck \cite{Diss_Steinbeck} as well as Saxena \& Joshi \cite{Saxena1989} revealed that $W$ creates tungsten oxide as well as tungsten nitride under the exposure of $N_{2}$ \& $O_{2}$. Less influence of adsorption during the calibration was observed with {platinum ($Pt$)} confirming the calibration of the measurement device (not shown).

%

\subsection{Molecular dynamics}
\begin{figure}[htpb]
\centering
\fontsize{9}{11}\selectfont 
 \input{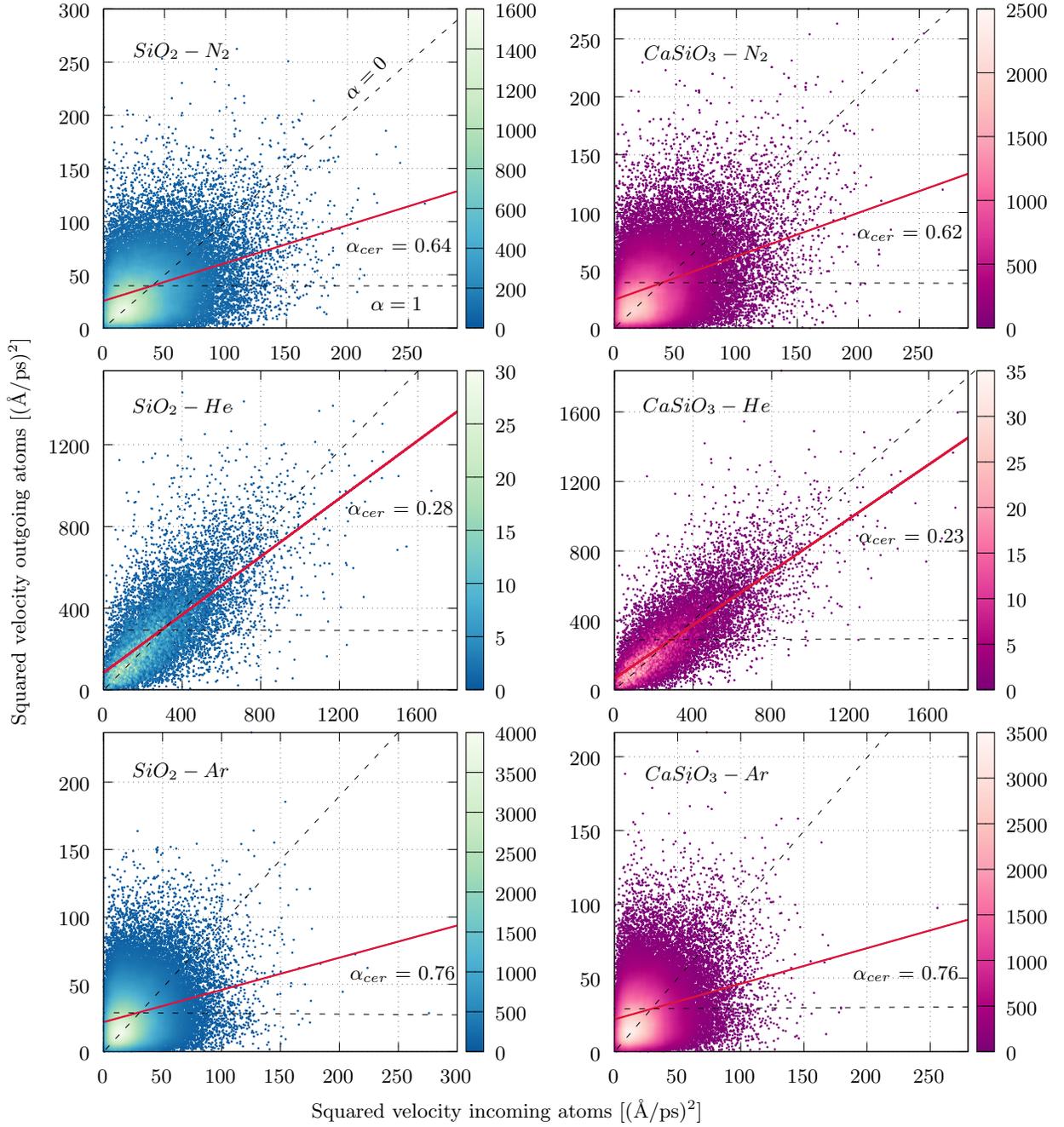}
\caption{Overview of final results from MD for $SiO_2$ (blue) and $CaSiO_3$ (pink) with Ar, He and N\textsubscript{2}; squared velocity of incoming atoms vs. squared velocity of outgoing atoms; diagonal limit (dashed line) $\rightarrow$ $\alpha_{cer}$ = 0, horizontal limit (dashed line) $\rightarrow$ $\alpha_{cer}$ = 1, red line = result of $\alpha_{cer}$ due to model after Spijker \cite{spijker2010computation} (least squares approximation); palette gives points density}
\label{fig:TACs}
\end{figure}

Results of MD for amorphous $SiO_2$ and $CaSiO_3$ are presented in Figure \ref{fig:TACs}. According to the model of Spijker \cite{spijker2010computation} the velocities of the incoming and outgoing atoms passing the virtual plane are tracked and used for {the} calculation of $\alpha_{cer}$ via a linear least squares approximation. By looking at the resulting plots {it might be} obvious that for smaller $\alpha_{cer}$ (e.g. $SiO_2$/$CaSiO_3$-He) the cluster of most points {has a more} elliptical {shape} than for {a} higher {value of} $\alpha_{cer}$ where it forms a circle. This is in accordance with Spijker \cite{spijker2010computation}. The limits of $\alpha_{cer}$ being 0 or 1 present {themselves} as dashed lines whereas the red line {represents} the final result of the linear least squares approximation. The slope of the red line gives $\alpha_{cer}$. Note that with an increasing slope $\alpha_{cer}$ decreases. Further, it appears that the results are quite equal for $SiO_2$ and $CaSiO_3$ for each gas. Only with decreasing atomic weight of the gas the deviations of $\alpha_{cer}$ between $SiO_2$ and $CaSiO_3$ increase significantly, but very slightly as with $CaSiO_3$ exhibiting the lower $\alpha_{cer}$. It is found that the effect of the type of thermostat used for controlling the temperatures of NVT regions on the TAC values, is insignificant and also initializing gas velocities with random seed numbers for the given temperature is found to have {a} negligible effect on $\alpha_{cer}$ due to the sufficient time allowed for the equilibration (see Figure \ref{fig:thermostats}).

\subsection{Experimental {R}esults}
\begin{figure}[H]
\centering
\fontsize{9}{11}\selectfont 
 \input{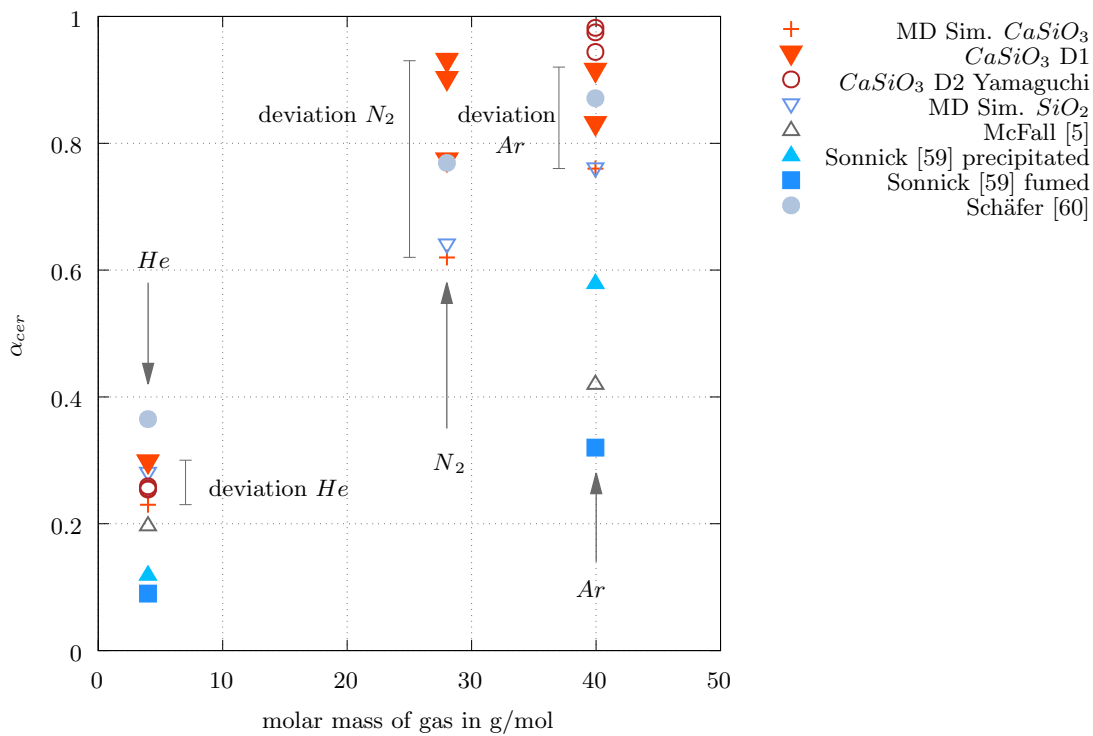}
\caption{Comparison of experimental (only $CaSiO_3$ D1, D2) and MD results ($SiO_2$ and $CaSiO_3$) as well as existing literature (only $SiO_2$) with $Ar, N_{2}$ and $He$ over molar mass of used gases; deviations between MD and exp. different for the gases; temperature ca. 345\,K for MD and exp., around 300\,K for literature, {standard uncertainty for own measurements with $He$ $\pm$28\,\% and with $Ar/N_{2}$ $\pm$5\% (rev 2 pt 3)}}
\label{fig:Messergebnisse-Ar}
\end{figure}

In Figure \ref{fig:Messergebnisse-Ar}, the final experimental results for $CaSiO_3$ D1 (smoothed surface) are shown versus the molar mass of the used gases in comparison with the results for $SiO_2$ and $CaSiO_3$ obtained from MD as well as from literature (only $SiO_2$). 
Several observations can be found from Figure \ref{fig:Messergebnisse-Ar}:
\begin{itemize}
 \item It {appears} that $\alpha_{cer}$ increases as the molar mass of the used gas increases. For $He$ the lowest values of $\alpha_{cer}$ were achieved as well as for $Ar$ the highest and with $N_{2}$ in between. This observation is valid for almost every result except for the experimental results where $\alpha_{cer}$ reaches an equal level for $N_{2}$ and $Ar$.

 \item The results from MD are lower than the experimental results for all gases. However, the deviations between experimental and MD results appear to be different being lowest for $He$, highest for $N_{2}$, whereas $Ar$ lies in between.
 
 \item Further promising results were obtained by Yamaguchi and Suzuki in a collaborative work, who used a spherical device for {the} measurement of $\alpha_{cer}$ (for description of the device see \cite{Yamaguchi2014}). Their results {resemble} the experimental data obtained during this {work}.
 \item In comparison to literature only those results of Schäfer \cite{Schaefer1954} agree with MD and the present experimental results. The data reported by McFall \cite{McFall1980} and Sonnick \cite{Sonnick2019} appear to be significantly lower for $He$ and even lower for $Ar$.
\item {An evaluation of the standard uncertainty of the experiments revealed very different values as the standard uncertainty for measurements with $He$ reaches up to $\pm$28\% whereas with $Ar$ and $N_2$ a standard uncertainty of $\pm$5\% is obtained. (rev 2 pt 3)}
\end{itemize}

\begin{figure}[H]
\centering
\fontsize{9}{11}\selectfont 
 \input{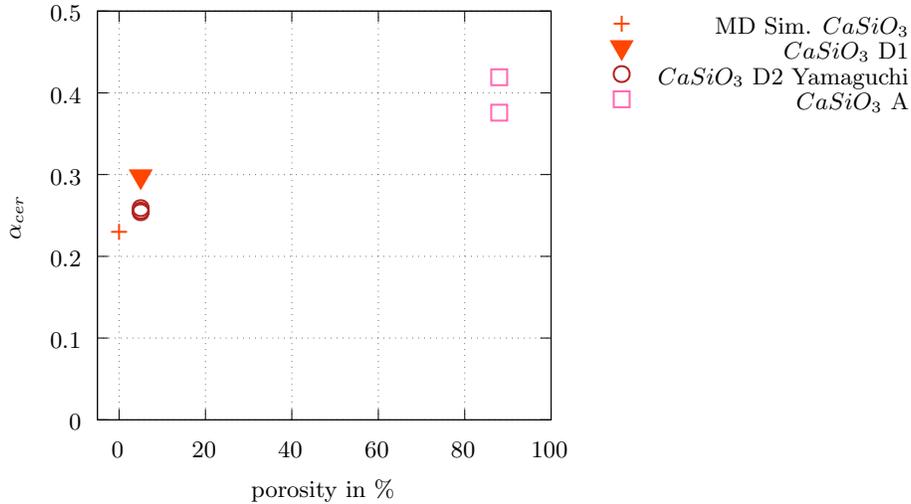}
\caption{Overview of experimental and simulation results of $CaSiO_3$ with $He$ over porosity; high porosity only for sample $CaSiO_3$ A available; temperature ca. 345\,K}
\label{fig:Messergebnisse-He}
\end{figure}

Another picture can be drawn from Figure \ref{fig:Messergebnisse-He}. It shows experimental results of $CaSiO_3$-He with different porosities (less 5\% D1, D2 and 88\% A). $\alpha_{cer}$ appears to be higher for higher porosity although no direct correlation could be made concerning the influence of roughness (see Table \ref{tab:Material}). Further, in this Figure the deviations between the results of MD (0\% porosity) and the experiment ($CaSiO_3$ D1 and D2 less than 5\% porosity) appear to show a very slight influence of porosity, whereas this is further discussed in the following section. Unfortunately, due to the difficulties in measurement and time constraints posed on the current work, further data is unavailable. We note, however, that this raises interesting questions for future research.

\newpage
\subsection{Discussion}
In the following a discussion of all obtained results is provided as several issues can be highlighted from Figures \ref{fig:TACs}, \ref{fig:Messergebnisse-Ar} and \ref{fig:Messergebnisse-He}:
\begin{enumerate}

\item A comparison between the experimental and molecular dynamic results shows a similar increasing trend versus the molar mass, although the values from MD are lower. This increasing trend is a well-known and expected behavior which was highlighted in literature (e.g. see Saxena \& Joshi \cite{Saxena1989}) and also in earlier simulation studies \cite{mohammad2020influence, liang2014parametric}. This point also demonstrates the validity of MD. Furthermore, the results of the MD can be used as a lower limit for $\alpha_{cer}$ on $SiO_2$ \& $CaSiO_3$ since they are determined on free surfaces, i.e. without taking adsorption effects and surface roughness into account. In the course of this ideal limit, based on the experimental results, an influence can be assumed since the deviations between the experimental and molecular-dynamic results are different. To discuss this further, one look back to Figure \ref{fig:W-Ar}: Zacharjin \cite{Zacharjin1936} carried out very interesting measurements by showing the influence of adsorption layers of contaminating gases. The author{s were} able to show that $Ar$, in the absence of external contamination, is desorbed significantly faster with increasing temperature (lower $\alpha_{W-Ar}$) than when a $N_{2}$ adsorption layer is present (higher $\alpha_{W-Ar}$). Accordingly, although $N_{2}$ has a lower molar mass than $Ar$, there is a higher enthalpy of adsorption for $N_{2}$ and thus a fundamentally higher thermal accommodation coefficient in general, which is also observed in the literature for other combinations of $N_{2}$ (e.g. near room temperature $W-N_{2}$: $\alpha_{W-N_2}$ $\approx$ 0.5 or $Pt-N_{2}$: $\alpha_{Pt-N_2}$ $\approx$ 0.8, see Saxena \& Joshi \cite{Saxena1989}) as well as in everyday laboratory work \cite{V-Jakobi}. \\
A similar behavior can be seen for the experimental results on $CaSiO_3$ D1. Regarding the MD simulation for $CaSiO_3$ D1-$Ar$ a higher value for $\alpha_{cer}$ can be expected due to the higher molar mass of $Ar$ than for $CaSiO_3$ D1-$N_2$. Nevertheless, the values for $CaSiO_3$ D1-$N_2$ are on a similar level as for $CaSiO_3$ D1-$Ar$ representing a significantly greater influence of an $N_{2}$ adsorption layer.

\item The experimental results of Yamaguchi and Suzuki ($CaSiO_3$ D2) agree very well with experimental results ($CaSiO_3$ D1). Furthermore, their results continue to exhibit similar deviations in comparison to MD which could be attributed to adsorption. Therefore, it empowers the assumption of an adsorptive influence on the surface and confirms our results.

\item The results of Sonnick et al. \cite{Sonnick2019} with precipitated and fumed silica are not of experimental nature. For {the} calculation of the thermal accommodation coefficient the model of Kaganer \cite{Kaganer1969} was used exclusively based on the molecular masses of gas and wall molecules. This model neglects any influences like temperature or adsorption or roughness, which can have major contributions to $\alpha$. {Further the experimental validation of Kaganers model fails as it cannot predict well-measured thermal accommodation coefficients, e.g. $\alpha_{W-Ar,exp}$ = 0.27 vs. $\alpha_{W-Ar,Kaganer}$ = 0.58 (at 300\,K). Therefore, the validity of the results for precipitated and fumed silica are questionable until the experimental proof. (rev 2 pt 1)}

\item {The data obtained by McFall \cite{McFall1980} are significantly lower. By examining the obtained data of McFall \cite{McFall1980} one can see that a steady state has not been reached by the end of the measurements as it was also stated in report of McFall \cite{McFall1980}. This opens the question of a continuous surface contamination which can have a huge impact on $\alpha$ as with increasing surface contamination the thermal accommodation coefficient increases, too \cite{Saxena1989}. (rev 2 pt 2)  }

 \item From MD the results of amorphous  $SiO_{2}$ (blue cross) and $CaSiO_{3}$ (red cross) reach very similar values for all examined gases (see Figure \ref{fig:Messergebnisse-Ar}). The reason for this close agreement can be attributed to their similar composition and certain similar short range order structural features such as the $SiO_{4}$ tetrahedron, and most importantly to the usage of the same gas-ion interaction potential parameter values for the dominant species {$Si$} \& {$O$} in both the cases. In addition, the gas-wall potential depth, which strongly influences values of $\alpha_{cer}$, was also comparable for both, the {$Ca$} and {$O$} ions.
 
 \item In MD, the slightly increasing difference of $\alpha_{cer}$ between $SiO_2$ and $CaSiO_3$ with decreasing molar mass of the gas, although the mass of the solid increases from $SiO_2$ to $CaSiO_3$, is attributed to the existence of $Ca$. As per the hard sphere model (which accounts only for direct specular collisions as observed in $He$), a large mismatch in molar mass between the gas ($Ar, N_{2}, He$) molecules and the solid ($Ca$) leads to inefficient energy transfer between them \cite{liang2014parametric}. Particularly for the case of $CaSiO_3$-$He$, the heavier $Ca$ (relative to the $He$ atom $\rightarrow$ mass of $He$ / mass of $Ca$ = 0.1) could be a potential reason for the reduction of $\alpha_{cer}$ as compared to $SiO_2$.

\item The last discussion point is for the measurement of $CaSiO_3$ A-He. The porosity of $CaSiO_3$ A is 88\%. Although the porosity and the surface roughness have no recognizable relationship, a higher $\alpha_{cer}$ is observed, which would be expected with a higher surface roughness (compare Table \ref{tab:Material}). One reason for this behavior could be due to the open porosity, where the gas particles also fly into the porous sample and collide very often with the solid structure resulting in an individual $\alpha$ close to 1. After a (longer) phase within the sample they return to the surface and fly into the free space of the gas gap. Since not every particle “dips” into the open-pored structure, it can be assumed that this effect only occurs in part. To investigate this effect further, the aim was to progress with the measurement of $CaSiO_3$ A-Ar. Unfortunately, however, it was not possible to determine the thermal accommodation coefficient because, as can be seen in Figure \ref{fig:Messfahigkeit}, there was a different trend of the measuring points (red crosses) over the Knudsen number in comparison to the model curve of $\lambda_{Gas}$ (blue line). A calculation of $\alpha_{cer}$ in this situation is not possible. $\lambda_{eff}$ for $CaSiO_3$ B (dotted line), which is very similar to $CaSiO_3$~A and was measured separately, increases in the range of measurability influencing the trend of $\lambda_{Gas}$ significantly, whereby the experimental values cannot be fitted to Springer's model.   



\begin{figure}[htpb]
 \centering
 \includegraphics[width=11cm]{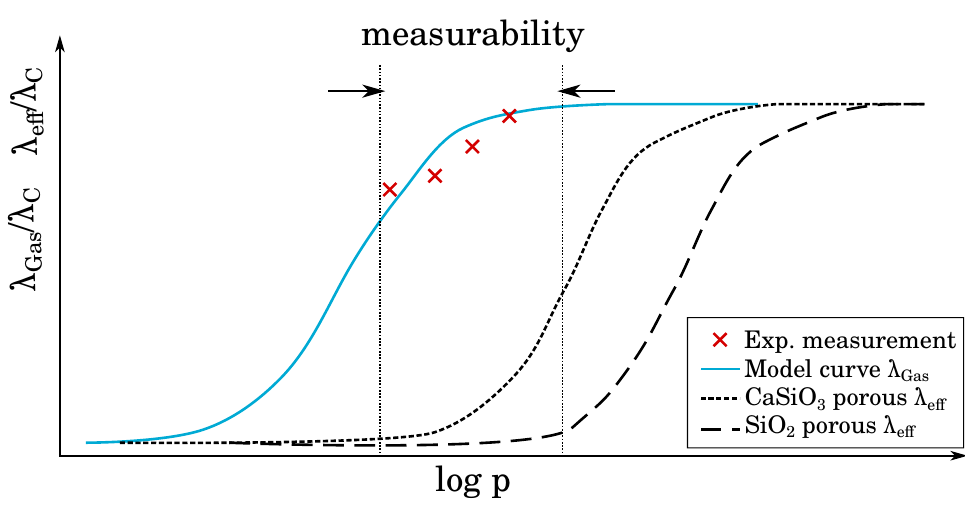}
 \caption{Scheme measurability of $CaSiO_3$ A with Ar, qualitive trends, thermal conductivity of gas and overall (effective) respectively over logarithmic pressure; blue line: model curve of gas thermal conductivity after Springer \cite{Springer1971}, red crosses: measurement points, dotted line:  effective thermal conductivity of highly porous $CaSiO_3$ B, dashed line: effective thermal conductivity of highly porous $SiO_2$  }
 \label{fig:Messfahigkeit}
\end{figure}

\end{enumerate}



\section{Conclusions}


A collaborative study between experiments and atomistic simulations on the determination of thermal accommodation coefficients on amorphous $CaSiO_3$ and $SiO_2$ with different gases is presented. The thermal accommodation coefficient $\alpha$, which is {amongst other applications} a part of the effective thermal conductivity of highly porous insulation materials, is usually set to unity for gases like $Ar$ and $N_{2}$ and 0.3 for $He$ {over a wider temperature range}, but {it} remained experimentally unverified {leaving the reliability of existing models of the effective thermal conductivity in question}. This open question was addressed using experimental measurements as well as molecular dynamics simulations and the results are in good agreement (with explainable deviations) with this widely used assumption. {It also confirms the validity of using this assumption in models of the effective thermal conductivity}. However, these results are only valid near room temperatures and molecular dynamics results showed no significant effect on $\alpha$ (only 1\% difference) by varying temperatures $\pm$ 20\,K around the room temperature, thus demanding no further investigation of this influence from the experimental side {in such short temperature range}. But from literature, significant influences on $\alpha$ at elevated and cryogenic temperatures are expected and could be an open topic for future research, especially using atomistic simulations. {Further}, though the influence of surface roughness, contamination and open porosities could not be examined robustly, due to insufficient measurement data to support clear scientific claims, yet some interesting observations are made which provide options to future research. 

\section{Acknowledgements}

Special thanks goes to Prof. Hiroki Yamaguchi and Mr. Yuta Suzuki from {the} University of Nagoya/Japan as they provided comparative measurements of $\alpha$ on $CaSiO_3$. Their work greatly supports the validity of the results of this work.\\
Further, great thanks goes to Dr. Claudia Voigt, who successfully prepared the sample $CaSiO_3$ D1, which {represents} a very important part of this {work}.\\
Computing time was {provided} by the high performance computing cluster (HPC) at TU Freiberg.\\
Funding: This work was supported by the German Research Foundation (DFG, GR1060/14) and a scholarship by the Saxon Ministry of Science and Art.

\section*{References}
\bibliography{DFG}

\end{document}